\documentstyle[12pt]{article}
\def\inprod#1#2{\left\langle #1, #2\right\rangle}
\newcommand{\rtimes}{\mbox{$\times\!\rule{0.3pt}{1.1ex}\,$}}
\def\ie{\hbox{\it i.e.}}
\def\eg{\hbox{\it e.g.}}
\def\bra#1{\left\langle #1\right |}
\def\ket#1{\left | #1\right\rangle}
\def\bracket#1#2{\left\langle #1 | #2 \right\rangle}
\def\vaca{\Omega_{\A}}
\def\vacu{\Omega_{\U}}
\def\A{{\cal A}}
\def\B{{\cal B}}
\def\C{{\cal C}}
\def\J{{\cal J}}
\def\P{{\cal P}}
\def\R{{\cal R}}
\def\U{{\cal U}}
\def\V{{\cal V}}
\def\idA{1_{\A}}
\def\idU{1_{\U}}
\def\idS{1_{\smash}}
\def\smash{{\A \rtimes \U}}
\def\id{\mathop{\rm id}}
\def\Y{\Upsilon}
\newcommand{\la}[1]{\stackrel{\leftarrow}{#1}}
\newcommand{\ra}[1]{\stackrel{\rightarrow}{#1}}
\begin{document}
\begin{titlepage}
\begin{center}
October 15, 1993	\hfill    LBL-33274 \\
			\hfill    UCB-PTH-92/42 \\

\vskip .5in

{\large \bf The Role of the Canonical Element in the Quantized Algebra of
Differential Operators $\smash$}
\footnote{This work was supported in part by the Director, Office of
Energy Research, Office of High Energy and Nuclear Physics, Division of
High Energy Physics of the U.S. Department of Energy under Contract
DE-AC03-76SF00098 and in part by the National Science Foundation under
grant PHY-90-21139.}

\vskip .5in

Chryssomalis Chryssomalakos, Peter Schupp and Paul
Watts\footnote{chryss@physics.berkeley.edu,
schupp@physics.berkeley.edu, watts@lbl.gov}

\vskip .5in

{\it  Theoretical Physics Group\\
      Lawrence Berkeley Laboratory\\
      University of California\\
      Berkeley, California 94720}
\end{center}

\vskip .5in

\begin{abstract}

We review the construction of the cross product algebra $\smash$ from
two dually paired Hopf algebras $\U$ and $\A$.  The canonical element
in $\U \otimes \A$ is then introduced, and its properties examined.
We find that it is useful for giving coactions on $\smash$, and it
allows the construction of objects with specific invariance properties
under these coactions.  A ``vacuum operator'' is found which projects
elements of $\smash$ onto said objects.  We then discuss bicovariant
vector fields in the context of the canonical element.

\end{abstract}
\end{titlepage}

\newpage
\renewcommand{\thepage}{\arabic{page}}
\setcounter{page}{1}

\section{Introduction}

Given two dually paired Hopf algebras $\U$ and $\A$, it is well known
that a new algebra, the cross product algebra $\smash$, can be
constructed, and that this algebra may be viewed as consisting of
differential operators and the functions they act on.  In Section 2,
we review the standard way of constructing this algebra
\cite{Sw,M,SWZ}.  The interpretation of this algebra as one of
{\em bicovariant} objects requires the introduction of well-defined
actions and coactions on $\smash$; these are given in Section 3, again
in the standard fashion \cite{SWZ}.

However, the emphasis of this paper will be on exploring the
properties of the canonical element $C$ of $\U \otimes \A$, which
enters very naturally through the expressions for the coactions, as is
shown in Section 4.  Furthermore, we can recover many of the familiar
relations for quantum groups from the consistency relations which $C$
satisfies in the case where $\U$ is quasitriangular \cite{RTF,Z}.  In
Section 5, we find new right and left invariant elements of $\smash$,
as well as an element which realizes the vacuum operator in $\smash$.
Viewed geometrically, this object projects vector fields and functions
onto their value at the identity.  Finally, in Section 6, we present a
method for constructing bicovariant vector fields in $\U$ with
particularly simple coactions, and examine its relation to the
canonical element.

Much of the notation used in this paper may be found in \cite{M,RTF}
and references therein.

\section{Hopf Algebras and their Duals, and the Cross Product}

We begin with a review of the basic relations for a Hopf algebra, and
the process by which its dual is also made a Hopf algebra.

Let $\U$ be a Hopf algebra, $\ie$ $\U$ is an associative algebra over
a field $k$ with unit $\idU$, and the coproduct $\Delta : \U
\rightarrow \U \otimes \U$, antipode $S:\U \rightarrow \U$, and counit
$\epsilon : \U \rightarrow k$ are all linear maps satisfying the usual
relations \cite{M}.  Let $\A$ be the dual of $\U$, $\ie$ the set of
linear functionals of elements of $\U$.  $\A$ is a vector space in the
usual manner, and can be endowed with a Hopf algebra structure via the
definitions
\begin{eqnarray}
\inprod{x}{ab}&\equiv &\inprod{\Delta(x)}{a\otimes b},\nonumber \\
\inprod{x}{\idA}&\equiv &\epsilon (x),\nonumber \\
\inprod{x\otimes y}{\Delta (a)}&\equiv &\inprod{xy}{a},\nonumber \\
\inprod{x}{S(a)}&\equiv &\inprod{S(x)}{a},\nonumber \\
\epsilon(a)&\equiv &\inprod{\idU}{a}, \label{dualreln}
\end{eqnarray}
($x,y \in \U$, $a,b \in \A$) which give, respectively, product, unit,
coproduct, antipode, and counit on $\A$.  $\U$ and $\A$ are said to be
{\em dually paired}.  We may now introduce a linearly independent
basis $\{ e_i \}$ for $\U$, $\ie$ $\U = \mbox{span}\{ e_i|i \in \J \}$
where $\J$ is the appropriate index set, and take the basis of $\A$ to
be $\{ f^i \}$ given by
\begin{equation}
\inprod{e_i}{f^j}=\delta_{i}^{j}.
\end{equation}
We will use these bases extensively throughout the remainder of this
paper.

Now that we have two dually paired Hopf algebras, we can introduce a
unital algebra which is denoted $\smash$, the ``cross product'' of
$\A$ and $\U$.  (This construction is also called the ``smash
product'', and is a Hopf algebra generalization of the Heisenberg
double and the Weyl semidirect product.)  $\smash$ is constructed to
be isomorphic to $\A \otimes \U$ as a vector space.  This may be seen
explicitly through the definition of the multiplication on $\smash$:
\begin{eqnarray}
a \cdot b &\equiv & ab\otimes \idU ,\nonumber \\
x \cdot y &\equiv & \idA \otimes xy, \nonumber \\
a \cdot x &\equiv & a\otimes x, \nonumber \\
x \cdot a &\equiv & a_{(1)}\otimes x_{(2)}\inprod{x_{(1)}}{a_{(2)}},
\label{mult}
\end{eqnarray}
where $a,b\in \A$, $x,y\in \U$, and we use the notation introduced by
Sweedler~\cite{Sw}:
\begin{equation}
\Delta (\alpha) = \alpha_{(1)} \otimes \alpha_{(2)}.
\end{equation}
Note that this multiplication is associative, and also that $\smash$
contains subalgebras isomorphic to both $\A$ and $\U$.  Therefore, we
shall drop the ``$\cdot$'' from now on, assuming that the above
multiplication is used when we take the product of an element of $\A$
with an element of $\U$.  Furthermore, we shall also write $e_i$ and
$f^i$ rather than $\idA \otimes e_i$ and $f^i \otimes \idU$
respectively when viewing the bases of $\U$ and $\A$ as elements in
$\smash$.

Notice that although $\U$ and $\A$ are both Hopf algebras, $\smash$ is
not, $\ie$ $\smash$ is an algebra that does not admit a coproduct or
counit (and therefore no antipode) even though the subalgebras $\U$
and $\A$ do.  Despite this fact, the original Hopf algebraic
structures of $\U$ and $\A$ are preserved.

\section{Actions and Coactions}
\label{sec-actions}
Suppose we have an algebra $\B$ and a vector space $\V$; a {\em left
action} of $\B$ on $\V$ is a bilinear map $\triangleright : \B \otimes
\V \rightarrow \V$ satisfying
\begin{equation}
(xy)\triangleright v =x\triangleright (y \triangleright v)
\end{equation}
for all $x,y \in \B$ and $v\in \V$.  A right action $\triangleleft$ of
$\B$ on $\V$ can be defined similarly.  In the case where $\B$ is a
Hopf algebra and $\V$ is a unital algebra, we further require that for
$x \in \B$ and $a,b \in
\V$,
\begin{eqnarray}
x \triangleright (ab) &=& (x_{(1)}\triangleright a)(x_{(2)}\triangleright b),
\nonumber \\
1_{\B} \triangleright a &=&a\nonumber\\
x \triangleright 1_{\V} &=& 1_{\V}\epsilon (x).
\end{eqnarray}
In this case, $\triangleright$ is called a {\em generalized
derivation}, and we can interpret $\B$ as an algebra of differential
operators which act on functions ($\ie$ elements of $\V$).

Now, suppose we have a coalgebra $\C$ and a vector space $\V$; a {\em
left coaction} of $\C$ on $\V$ is a linear map $_{\C} \Delta: \V
\rightarrow \C \otimes \V$ satisfying
\begin{eqnarray}
(\Delta \otimes \id)_{\C}\Delta(v)&=&(\id \otimes {}_{\C}\Delta)_{\C}
\Delta (v),\nonumber \\
(\epsilon \otimes \id)_{\C}\Delta(v)&=&v,
\end{eqnarray}
for all $v \in \V$, where $\Delta$ and $\epsilon$ are the coproduct
and counit on $\C$, respectively.  The right coaction $\Delta_{\C}$ is
defined similarly.  If $\C$ is a Hopf algebra and $\V$ is a unital
algebra, we impose the further conditions that
\begin{eqnarray}
_{\C}\Delta (ab)&=&{}_{\C}\Delta (a)_{\C}\Delta (b),\nonumber \\
_{\C}\Delta(1_{\V})&=&1_{\C}\otimes 1_{\V},
\end{eqnarray}
for $a, b \in \V$, $\ie$ $_{\C}\Delta$ must be an algebra
homomorphism.

We now introduce specific actions and coactions in the case where we
have the two Hopf algebras $\U$ and $\A$ and the associative unital
algebra $\smash$.  The left and right actions of $\U$ on $\smash$ are
defined as
\begin{eqnarray}
x\triangleright \alpha &\equiv & x_{(1)}\alpha S(x_{(2)}), \nonumber \\
\alpha \triangleleft x &\equiv & S(x_{(1)})\alpha x_{(2)},
\end{eqnarray}
for $x \in \U$, $\alpha \in \smash$.  Note that for the case where
$\alpha \in \U$, these are the left and right adjoint actions, and
when $\alpha \in \A$, the right action of $x$ on $\alpha$ may be
written
\begin{equation}
x \triangleright \alpha = \alpha_{(1)} \inprod{x}{\alpha_{(2)}},\label{diffact}
\end{equation}
which is the usual action of a differential operator $x$ on a function
$\alpha$.

Keeping in mind that the coaction should describe the transformation
properties of the elements of $\smash$, we make the following choices:
$\A$ left coacts on $\smash$ so as to leave $\U$ invariant, $\ie$
\begin{equation}
_{\A}\Delta (x) \equiv \idA \otimes x,
\end{equation}
$x \in \U$.  Furthermore, $\A$ left and right coacts on $\A$ via
the coproduct:
\begin{equation}
_{\A}\Delta (a) = \Delta_{\A} (a)= \Delta (a), \label{coprod}
\end{equation}
$a \in \A$.  The right coaction of $\A$ on $\U$ is a bit more
complicated; we first use a Sweedler-like notation to write the
coactions as
\begin{eqnarray}
_{\A}\Delta (\alpha)\equiv \alpha^{(1')}\otimes \alpha^{(2)},& &
\Delta_{\A} (\alpha)\equiv \alpha^{(1)} \otimes \alpha^{(2')},
\end{eqnarray}
where the unprimed elements live in $\smash$, the primed ones in $\A$.
Inspired by the form of (\ref{diffact}), we therefore define the two
pieces of $\Delta_{\A}(x)$ for $x \in \U$ to be those quantities which
satisfy
\begin{equation}
y \triangleright x \equiv x^{(1)} \inprod{y}{x^{(2')}}
\end{equation}
for $y\in \U$.  This may look a bit mysterious, and one might wonder
if this really defines both $x^{(1)}$ and $x^{(2')}$.  To see that it
actually does, we use our dual basis to write $\Delta_{\A}(x)$ as
\begin{equation}
\Delta_{\A}(x) \equiv x_i \otimes f^i ,
\end{equation}
where $x_i \in \U$.  (Throughout this paper, repeated indices are
summed over the index set $\J$.)  Therefore,
\begin{equation}
e_j \triangleright x = x_i \inprod{e_j }{f^i } = x_j ,
\end{equation}
giving
\begin{equation}
\Delta_{\A}(x)=(e_i \triangleright x)\otimes f^i .\label{coact}
\end{equation}
All of the above definitions are consistent with the conditions
necessary for $\Delta_{\A}$ to be a right coaction.

Similarly, we can define a left coaction of $\U$ on $\smash$:
\begin{equation}
_{\U}\Delta : \smash \rightarrow \U \otimes \smash ;  \; \; \; \smash \ni
\alpha \mapsto {}_{\U}\Delta (\alpha ) \equiv \alpha^{(\bar{1} )} \otimes
\alpha^{(2)}.
\end{equation}
On $\U$, $_{\U}\Delta$ is the coproduct:
\begin{equation}
_{\U}\Delta (x) \equiv x^{(\bar{1} )} \otimes x^{(2)}=x_{(1)} \otimes x_{(2)}
\equiv \Delta (x), \; \; \; x \in \U .
\end{equation}
On  $\A$, $_{\U}\Delta$ is defined again implicitly via
\begin{equation}
ab=b_{(1)} \langle a^{(\bar{1} )} ,b_{(2)} \rangle a^{(2)}, \; \; \; \; a,b
\in \A .
\end{equation}
Using the right action of a function $b$ on another function $a$ given by
\begin{equation}
a \triangleleft b \equiv S(b_{(1)}) a b_{(2)},
\end{equation}
one can easily show that
\begin{equation}
_{\U}\Delta (a) = e_i \otimes (a \triangleleft f^i ).
\end{equation}

In the following sections we focus on the nontrivial coactions $\Delta
_{\A}$ and $_{\U}\Delta$ which, for simplicity, we refer to as the
right and left coactions respectively. For example, an element
$\alpha$ of $\smash$ will be called ``left invariant'' if $_{\U}\Delta
(\alpha )=\idU \otimes \alpha$, while ``right invariant'' elements
satisfy $\Delta _{\A} (\alpha )=\alpha \otimes \idA$.

\section{The Canonical Element}

We now introduce the canonical element $C$, which lives in $\U \otimes \A$, and
has the form
\begin{equation}
C\equiv e_i \otimes f^i.
\end{equation}
$C$ satisfies several relations; for instance, note that
\begin{eqnarray}
(\Delta \otimes \id)(C)&=& \Delta (e_i )\otimes f^i \nonumber \\
&=& (e_i )_{(1)}\otimes (e_i )_{(2)}\otimes f^i \nonumber \\
&=& e_i \otimes e_j \otimes f^i f^j \nonumber \\
&=& (e_i \otimes \idU \otimes f^i )(\idU \otimes e_j \otimes f^j )
\nonumber\\
&=& C_{13}C_{23}
\end{eqnarray}
(where in going from the second to the third line we have used the
duality between $\U$-comultiplication and $\A$-multiplication).
Similar calculations also give $(\id \otimes \Delta)(C)=C_{12}C_{13}$,
as well as the following:
\begin{eqnarray}
(S \otimes \id)(C)=(\id \otimes S)(C)&=&C^{-1},\nonumber \\
(\epsilon \otimes \id)(C)=(\id \otimes \epsilon)(C)&=&\idU \otimes \idA.
\end{eqnarray}
$C$ does more than just satisfying the above relations; to see that
this is true, we can compute the right coaction of a basis vector in
$\U$.  Using (\ref{coact}),
\begin{eqnarray}
\Delta_{\A}(e_i )&=&(e_j \triangleright e_i )\otimes f^j \nonumber \\
 &=& (e_j )_{(1)}e_i S((e_j )_{(2)})\otimes f^j \nonumber \\
 &=& e_m e_i S(e_n )\otimes f^m f^n \nonumber \\
 &=& (e_m \otimes f^m )(e_i \otimes \idA)(S(e_n )\otimes f^n )\nonumber \\
 &=& C(e_i \otimes \idA)(S \otimes \id)(C),
\end{eqnarray}
so for any $x \in \U$,
\begin{equation}
\Delta_{\A}(x)=C(x\otimes \idA)C^{-1}.
\end{equation}
However, when we think of $C$ as living in $(\smash)\otimes (\smash)$,
with $e_i$ and $f^i$ as the bases for the subalgebras $\U$ and $\A$ of
$\smash$ respectively, further results follow; for instance, if $a \in
\A$,
\begin{eqnarray}
C(a\otimes \idS)C^{-1}&=&e_i aS(e_j )\otimes f^i f^j \nonumber \\
&=&(a_{(1)}(e_i )_{(2)}\inprod{(e_i )_{(1)}}{a_{(2)}})S(e_j )\otimes
f^i f^j \nonumber \\
&=&a_{(1)}\inprod{(e_k )_{(1)}}{a_{(2)}}(e_k )_{(2)}S((e_k )_{(3)})\otimes
f^k \nonumber \\
&=&a_{(1)}\otimes \inprod{e_k }{a_{(2)}}f^k \nonumber \\
&=&a_{(1)}\otimes a_{(2)},
\end{eqnarray}
(where $\idS\equiv \idA \otimes \idU$) so that
\begin{equation}
C(a\otimes \idS)C^{-1}=\Delta (a).
\end{equation}
Thus, the right coaction of $\A$ on $\smash$ can be written as
\begin{equation}
\Delta_{\A}(\alpha )=C(\alpha \otimes \idS)C^{-1} \label{gencoact}
\end{equation}
for any $\alpha \in \smash$.  (This expression shows explicitly that
$\Delta_{\A}$ is an algebra homomorphism.)  We can continue doing
calculations along these lines, and we find
\begin{equation}
C^{-1}(\idS \otimes \alpha )C= {}_{\U}\Delta (\alpha)
\end{equation}
for $\alpha \in \smash$ (so that, for $x \in \U$, $\Delta
(x)=C^{-1}(\idS \otimes x)C)$.  Using these results, together with the
coproduct relations for $C$, we obtain the equation
\begin{equation}
C_{23}C_{12}=C_{12}C_{13}C_{23}.\label{YBE}
\end{equation}
(Interestingly, this equation can be viewed as giving the multiplication on
$\smash$ as defined in (\ref{mult}).)

In the case where $\U$ is a quasitriangular Hopf algebra with universal
R-matrix $\R$, the coproduct relations involving $C$ imply the following
consistency conditions:
\begin{eqnarray}
\R_{12}C_{13}C_{23}&=&C_{23}C_{13}\R_{12},\nonumber \\
\R_{23}C_{12}&=&C_{12}\R_{13}\R_{23}, \nonumber \\
\R_{13}C_{23}&=&C_{23}\R_{13}\R_{12}.\label{quasi}
\end{eqnarray}
To see the added significance of these equations, note that
\begin{equation}
\inprod{C}{a \otimes \id}=a,
\end{equation}
where $a \in \A$, and we use the convention
\begin{equation}
\inprod{\alpha }{\id}=\alpha
\end{equation}
for any $\alpha\in\smash$.  Let $\rho : \U \rightarrow M_{n}(k)$ be an
$n\times n$ matrix representation of $\U$, and define the $n^2$ matrix
elements $A^{i}{}_{j}\in {\cal A}$ by
\begin{equation}
\inprod{x}{A^{i}{}_{j}}\equiv \rho ^{i}{}_{j}(x).
\end{equation}
(These $A^{i}{}_{j}$s are what are usually viewed as the noncommuting
matrix elements of the pseudomatrix group associated with $\U$
\cite{W}.)  Given $\rho$, we can define the $\U$-valued matrices
$L^{\pm}$ by
\begin{eqnarray}
L^{+}&\equiv &(\id \otimes \rho)(\R),\nonumber \\
L^{-}&\equiv &(\rho \otimes \id)(\R^{-1}),
\end{eqnarray}
and the numerical R-matrix by
\begin{equation}
R\equiv (\rho \otimes \rho)(\R).
\end{equation}
Furthermore, it is easily seen that $(\rho \otimes \id)(C)=A$.  Now
let us apply $(\rho^{i}{}_{k}\otimes \rho^{j}{}_{l}\otimes \id)$ to
the first of equations (\ref{quasi}); the left hand side gives
\begin{eqnarray}
(\rho^{i}{}_{k}\otimes \rho^{j}{}_{l} \otimes \id)(\R_{12}C_{13}C_{23})&=&
(\rho^{i}{}_{m}\otimes \rho^{j}{}_{n})(\R)(\rho^{m}{}_{k}\otimes \id)(C)
(\rho^{n}{}_{l}\otimes \id)(C)\nonumber \\
&=&R^{ij}{}_{mn}A^{m}{}_{k}A^{n}{}_{l}.
\end{eqnarray}
The right hand side gives $A^{i}{}_{m}A^{j}{}_{n}R^{mn}{}_{kl}$, so
using the usual notation, we obtain
\begin{equation}
RA_{1}A_{2}=A_{2}A_{1}R,
\end{equation}
which gives the commutation relations between the elements of $A$.
Doing similar gymnastics with the other two equations in (\ref{quasi})
gives
\begin{eqnarray}
L^{+}_{1}A_{2}&=&A_{2}R_{21}L^{+}_{1},\nonumber \\
L^{-}_{1}A_{2}&=&A_{2}R^{-1}L^{-}_{1},
\end{eqnarray}
which give the commutation relations between elements of $\U$ and $\A$
within $\smash$.  (Of course, we also have the commutation relations
\begin{eqnarray}
RL_{2}^{\pm}L_{1}^{\pm}&=&L_{1}^{\pm}L_{2}^{\pm}R,\nonumber \\
RL_{2}^{+}L_{1}^{-}&=&L_{1}^{-}L_{2}^{+}R,
\end{eqnarray}
between elements of $\U$, obtained as above from $\R_{12} \R_{13}
\R_{23}=\R_{23} \R_{13} \R_{12}$, the quantum Yang-Baxter equation.)
Thus, we recover all the commutation relations between $A$ and
$L^{\pm}$ given in \cite{Z}.  In this case, $\A$ is interpreted as the
function algebra on the quantum group generated by the elements of
$A$, and $\U$ as the universal enveloping algebra generated by
elements of $L^{\pm}$.

\section{The Vacuum Operator}

An arbitrary element $\alpha$ of $\smash$ will, in general, have
nontrivial left and right coactions (as defined at the end of Section
\ref{sec-actions}). We can, however, associate to each such element
one left invariant and one right invariant element of $\smash$
(denoted $\stackrel{\rightarrow}{\alpha}$ and
$\stackrel{\leftarrow}{\alpha}$ respectively), given by the formulae
\begin{equation}
\stackrel{\rightarrow}{\alpha}=\alpha ^{(2)} S^{-1}(\alpha ^{(\bar{1} )} ),
\; \; \; \; \; \stackrel{\leftarrow}{\alpha}=S^{-1}(\alpha ^{(2')})
\alpha ^{(1)}.
\end{equation}
One may easily check that $_{\U}\Delta(\stackrel{\rightarrow}{\alpha})
=\idU\otimes \stackrel{\rightarrow}{\alpha}$ and $\Delta_{\A}
(\stackrel{\leftarrow}{\alpha})=\stackrel{\leftarrow}{\alpha} \otimes
\idA$ for $\alpha \in \smash$.

Notice that for $x \in \U$,
\begin{equation}
\stackrel{\rightarrow}{x}=x^{(2)}S^{-1}(x^{(\bar{1})})=x_{(2)}S^{-1}(x_{(1)})=
\epsilon (x) \idU,  \label{eq-rax}
\end{equation}
and similarly, for $a \in \A$,
\begin{equation}
\label{eq-laa}
\stackrel{\leftarrow}{a}=S^{-1}(a_{(2)}) a_{(1)}= \epsilon (a) \idA .
\end{equation}
Recall now the definitions of left and right vacua \cite{Z}, denoted
by $\vaca$ and $\vacu$ respectively, which satisfy
\begin{eqnarray}
L^{+} \vacu =L^{-} \vacu = I \vacu,&&\vaca A= \vaca I,\nonumber\\
\bracket{\vaca}{\vacu}&=&1,\label{vacuum}
\end{eqnarray}
where $I$ is the unit matrix. These have, so far, been introduced ``by
hand''.  We now show that they can be related to a special element of
$\smash$.  Consider the element $E$ of $\smash$, constructed out of
$C$ as follows:
\begin{equation}
E= m_{\smash} \circ (S^{-1}\otimes \id) \circ \tau (C) =S^{-1}(f^i )e_i
\end{equation}
(with $m_{\smash}$ the multiplication on $\smash$ and $\tau (\alpha \otimes
\beta)=\beta \otimes \alpha$ for all $\alpha,\;\beta \in \smash$). We
easily find that $E^{2}=E$. Furthermore, for $a \in \A$,
\begin{eqnarray}
Ea & = & S^{-1}(f^i )e_i a \nonumber \\
   & = & S^{-1}(f^i )a_{(1)} \inprod{(e_i)_{(1)}}{a_{(2)}}(e_i)_{(2)}
\nonumber \\
   & = & S^{-1}(f^i f^j )a_{(1)}\inprod{e_i }{a_{(2)}}e_j  \nonumber \\
   & = & S^{-1}(f^j )S^{-1}(a_{(2)})a_{(1)}e_j  \nonumber \\
   & = & E \epsilon (a), \label{eq-Ea}
\end{eqnarray}
and similarly, for $x \in \U$,
\begin{equation}
xE=\epsilon (x)E.\label{eq-xE}
\end{equation}
Note that (\ref{eq-Ea}) and (\ref{eq-xE}) imply that
\begin{equation}
ExaE=\inprod{x}{a}E \mbox{ and } EaxE=\epsilon(x)\epsilon(a)E.
\end{equation}
Provided we generalize (\ref{vacuum}) so that for $x\in \U$, $a\in
\A$,
\begin{eqnarray}
x\vacu = \vacu x &=&\epsilon(x)\vacu,\nonumber \\
\vaca a =a \vaca &=&\vaca \epsilon(a),
\end{eqnarray}
(which give as a consequence $\bracket{\vaca}{xa\,\vacu}=
\inprod{x}{a}$), the preceding properties of $E$ suggest a
representation given by
\begin{equation}
E \simeq \ket{\vacu}\bra{\vaca},
\end{equation}
so that ``vacuum operator'' is perhaps an appropriate name for $E$.

\noindent {\bf Remark:}  There also exists an object $\bar{E}$, given by
\begin{equation}
\bar{E}=m_{\smash}\circ (S^2 \otimes \id)(C)\equiv S^2 (e_i)f^i
\end{equation}
which has properties similar to that of $E$, $\eg$ $\bar{E}^2 =
\bar{E}$, $\bar{E} x=\bar{E} \epsilon(x)$ and $a \bar{E}
=\epsilon(a)\bar{E}$ for $x\in \U$, $a\in \A$; thus, $\bar{E}\simeq
\ket{\vaca}\bra{\vacu}$ is the vacuum operator ``adjoint'' to $E$.

The element $E$, however, has more properties. For example, for $x \in
\U$,
\begin{eqnarray}
E \la{x} & = & E S^{-1}(x^{(2')})x^{(1)} \nonumber \\
         & = & E \epsilon (x^{(2')}) x^{(1)} \nonumber \\
         & = & E x
\end{eqnarray}
(where we have used $(\id \otimes \epsilon)\circ \Delta_{\A}=\id$).
Similarly, we find for $a \in \A$,
\begin{equation}
a E=\ra{a}E.
\end{equation}
With the help of (\ref{eq-rax}) and (\ref{eq-laa}), we can summarize:
\begin{equation}
\alpha E=\ra{\alpha}E, \; \; \; \; E\alpha =E\la{\alpha}, \; \;
\; \; \alpha \in \smash .
\end{equation}

It is interesting to note that, in the classical limit, $E$ becomes a
formal Taylor expansion operator; $\eg$ $E$ applied to a function
$f(x)$ on a one-dimensional space returns $f(0)=f(x-x)$ \cite{BZpc}.
A more detailed exposition of these matters will appear in a
forthcoming paper \cite{CS}.

\section{Bicovariant Vector Fields}

The appearance of a (possibly) infinite sum in equation (\ref{coact}),
or for that matter (\ref{gencoact}), suggests that the elements of
$\U$ have in general very complicated transformation properties. In
contrast, the elements of $\A$, especially those constructed from the
matrix entries of $A$, have very simple transformation properties
given by the coproduct in $\A$ (\ref{coprod}).  We would like to show
how to construct vector fields corresponding to --- and inheriting the
simple behavior of --- these functions.  This construction can then be
used to find a basis for vector fields that closes under coaction and
hence under (mutual) adjoint actions.  First we need to prove the
following lemma:

\noindent {\bf Lemma: } {\em Let $\Y \equiv \Y_i \otimes \Y^i \in \U
\otimes \U$ be such that $\Y \Delta(x) = \Delta(x) \Y$ for all $x \in
\U$; it then follows that $\: \Y_i \otimes (x \triangleright \Y^i) =
(\Y_i \triangleleft x)\otimes \Y^i$ with $\Y_i \triangleleft x \equiv
S(x_{(1)}) \Y_i x_{(2)}$ for all $x \in \U$.}

\noindent {\bf Proof:}
\begin{equation}
\begin{array}{rcl}
\Y_i \otimes (x \triangleright \Y^i) & \equiv &
             \Y_i \otimes x_{(1)} \Y^i S(x_{(2)})\\
& = & S(x_{(1)}) x_{(2)} \Y_i \otimes x_{(3)} \Y^i S(x_{(4)})\\
& = & S(x_{(1)}) \Y_i x_{(2)} \otimes \Y^i x_{(3)} S(x_{(4)})\\
& = & (\Y_i \triangleleft x) \otimes \Y^i.\,\Box
\end{array}
\end{equation}
For any function $b \in \A$, define
\begin{equation}
Y_b \equiv \inprod{\Y}{b \otimes \id } \: \in \U .
\end{equation}
This vector field has the following transformation property:
\begin{equation}
\Delta_{\A}(Y_b) = Y_{b_{(2)}} \otimes S(b_{(1)}) b_{(3)}
\end{equation}
{\bf Proof:}
\begin{equation}
\begin{array}{rcl}
\Delta_{\A}(Y_b) & = & \inprod{\Y_i}{b} (e_k \triangleright \Y^i)
\otimes f^k\\
&=& \inprod{\Y_i \triangleleft e_k}{b} \Y^i \otimes f^k\\
&=& \inprod{\Y_i \otimes e_k}{b_{(2)}
    \otimes S(b_{(1)}) b_{(3)}} \Y^i \otimes f^k\\
& = & Y_{b_{(2)}} \otimes S(b_{(1)}) b_{(3)}.\,\Box
\end{array}
\end{equation}

\noindent {\bf Example:} Let $\Y \equiv \R_{21} \R_{12}$ and $b
\equiv A^i{}_j$; then $Y^i{}_j \equiv $$Y_{A^i{}_j} =$$\inprod{\R_{21}
\R_{12}} {A^i{}_j \otimes i\!d}$ is the well-known matrix of vector
fields $L^+ S(L^-)$ introduced in \cite{RS} with coaction
$\Delta_{\A}(Y^i{}_j) = Y^k{}_l \otimes S(A^i{}_k) A^l{}_j$.

This last example may in some cases provide a way of computing the
canonical element $C$ from $\R_{21} \R_{12}$:  let $\mu$ be the map
\begin{equation}
\mu:\A \rightarrow \U,\: b \mapsto \inprod{\R_{21}\R_{12}}{b \otimes \id}.
\end{equation}
There is a certain class of quasitriangular Hopf algebras for which
this map is invertible ($\ie$ the factorizable quasitriangular Hopf
algebras); therefore, if $\U$ is factorizable, the fact that
\begin{eqnarray}
(\id \otimes \mu)(C)&=&e_i \inprod{\R_{21} \R_{12}}{f^i \otimes
\id}\nonumber\\
&=&\R_{21}\R_{12}
\end{eqnarray}
implies that we can find an explicit form for $C$:
\begin{equation}
C = (\id \otimes \mu^{-1}) (\R_{21} \R_{12}).
\end{equation}

\section{Conclusion}

In this letter we have shown how the cross product algebra $\smash$
(interpreted as a quantized algebra of differential operators and
functions) constructed from the two dual Hopf algebras $\A$ and $\U$
admits transformations ($\ie$ coactions) of its elements given through
conjugation by the {\em canonical element} $C$ of $\U \otimes \A
\subseteq (\smash) \otimes (\smash)$.  We therefore have, in
principle, a way of finding the coactions on the differential
operators and functions, although the actual computations using $C$
may be difficult.  Since the coactions on $\smash$ can be expressed as
inner automorphisms, they are manifestly homomorphic, and they
explicitly preserve the commutation relations (\ref{mult}).

We have also found a general way of constructing objects in $\smash$
with particular invariance properties, $\eg$ elements $\ra{\alpha}$
which are invariant under the left coaction of $\U$.  In doing so, we
note the existence of the vacuum operator $E$, which projects out the
left or right invariant parts of elements of $\smash$.

The question of possibly finding $C$ more specifically than in terms
of the bases of $\U$ and $\A$ depends on the invertibility of the map
$\mu$; unfortunately, there does not seem to be any a priori reason to
assume that $\mu^{-1}$ exists in the nonfactorizable case \cite{Rpc},
so the general expression for $C$ may be as good as we can get.

As a final comment, we note that the canonical element is somewhat
reminiscent of the universal R-matrix constructed via the Drinfel'd
quantum double \cite{Drin}.  However, the similarity is only
superficial; for instance, $C$ does not satisfy a Yang-Baxter
equation, but instead satisfies (\ref{YBE}).  Even so, this similarity
suggests the possibility of axiomatizing the definition of a cross
product algebra in the same way as we define a quasitriangular Hopf
algebra, by postulating the existence of an associative unital algebra
$\P$, together with an element $C \in \P \otimes \P$ which satisfies
(\ref{YBE}).  A cross product algebra would then be a specific case of
this, in analogy to the fact that a quantum double is a particular
type of quasitriangular Hopf algebra.

\section*{Acknowledgment}

We would like to express our sincere gratitude to Professor B. Zumino
for advice and support and we would like to thank Professor N. Yu.
Reshetikhin for helpful discussions.

This work was supported in part by the Director, Office of Energy
Research, Office of High Energy and Nuclear Physics, Division of High
Energy Physics of the U.S. Department of Energy under Contract
DE-AC03-76SF00098 and in part by the National Science Foundation under
grant PHY-90-21139.

\end{document}